\documentclass[aip,11pt,reprint]{revtex4-1}

\usepackage{graphicx}   
\usepackage{color}      
\usepackage{hyperref} 
\usepackage[english]{babel}

\begin{document}

\title{Electro-optic routing of photons from single quantum dots in photonic integrated circuits}
\author{Leonardo Midolo}
\author{Sofie L. Hansen}
\affiliation{Niels Bohr Institute, University of Copenhagen, Blegdamsvej 17, DK-2100 Copenhagen, Denmark}
\author{Weili Zhang}
\affiliation{Niels Bohr Institute, University of Copenhagen, Blegdamsvej 17, DK-2100 Copenhagen, Denmark}
\affiliation{Key Laboratory of Optical Fiber Sensing and Communications (Education Ministry of China), University of Electronic Science and Technology of China, Chengdu, 611731, China}
\author{Camille Papon}
\affiliation{Niels Bohr Institute, University of Copenhagen, Blegdamsvej 17, DK-2100 Copenhagen, Denmark}
\author{R{\"u}diger Schott}
\author{Arne Ludwig}
\author{Andreas D. Wieck}
\affiliation{Lehrstuhl f\"ur Angewandte Festk\"orperphysik, Ruhr-Universit\"at Bochum, Universit\"atstrasse 150, D-44780 Bochum, Germany}
\author{Peter Lodahl}
\author{S{\o}ren Stobbe}
\affiliation{Niels Bohr Institute, University of Copenhagen, Blegdamsvej 17, DK-2100 Copenhagen, Denmark}


\begin{abstract}
Recent breakthroughs in solid-state photonic quantum technologies enable generating and detecting single photons with near-unity efficiency as required for a range of photonic quantum technologies. The lack of methods to simultaneously generate and control photons within the same chip, however, has formed a main obstacle to achieving efficient multi-qubit gates and to harness the advantages of chip-scale quantum photonics. Here we propose and demonstrate an integrated voltage-controlled phase shifter based on the electro-optic effect in suspended photonic waveguides with embedded quantum emitters. The phase control allows building a compact Mach-Zehnder interferometer with two orthogonal arms, taking advantage of the anisotropic electro-optic response in gallium arsenide. Photons emitted by single self-assembled quantum dots can be actively routed into the two outputs of the interferometer. These results, together with the observed sub-microsecond response time, constitute a significant step towards chip-scale single-photon-source de-multiplexing, fiber-loop boson sampling, and linear optical quantum computing.
\end{abstract}

\maketitle

\section{Introduction}
Using single photons as carriers of quantum information has become increasingly attractive due to the recent progress in integrated nanophotonic technology and solid-state devices, which allow generating and detecting photonic qubits within a chip\cite{obrien_photonic_2009}.
In particular, single-photon sources based on semiconductor quantum dots (QDs) and single-photon detectors based on superconducting thin films, have emerged as enabling technologies for on-chip quantum information processing with near-unity efficiency\cite{lodahl_interfacing_2015, Hofling_GaAs_integrated_2016, arcari_near-unity_2014, sprengers_waveguide_2011, marsili_detecting_2013}. 
An additional requirement for photonic quantum simulation, or even computation, is the availability of dynamically tunable optical elements. A central tunable element is a variable phase shifter, which can be used for building switches or tunable beam splitters. This is key to perform linear optical quantum computing \cite{knill_scheme_2001}, to recent proposals for boson-sampling\cite{wang_high-efficiency_2017,motes_scalable_2014} and to realize routers for de-multiplexed single-photon sources\cite{lenzini_active_2016}. While free-space Pockels cells or fiber-coupled electro-optic modulators are readily available, scalable quantum technology requires chip-scale routing of single-photons to retain high efficiency while scaling the number of circuit elements involved. 

Quantum dots in gallium arsenide (GaAs) nano-membranes constitute a well-established platform for combining high single-photon generation efficiency\cite{arcari_near-unity_2014} and indistinguishability\cite{somaschi_near-optimal_2016, Kirsanske_indistinguishable_2017} with planar waveguide devices. While this technology could be interfaced with various photonic circuits in multichip configurations, a scalable solution requires the integration of QDs with devices capable of dynamically routing single photons. This approach is radically different than previous works based on switching the emitter itself\cite{bentham_-chip_2015,luxmoore_optical_2013,sollner_deterministic_2015}. The integration of a broadband switch in a GaAs membrane for single-photon routing has been held back by the lack of a method to achieve large refractive-index modulation with fast response times at cryogenic temperatures.  
Approaches to on-chip routing of single photons using circuits have been reported, based on thermo-optical \cite{silverstone_-chip_2014}, electro-mechanical \cite{poot_broadband_2014}, and electro-optical\cite{Wang_GaAs_circuit_2014, jin_-chip_2014, sharapova_generation_2017} phase shifters. Yet, none of these works have shown routing of QD sources, let alone integration within the same chip. 
While the thermo-optic effect only works at room temperature and it is thus not compatible with QDs, the electro-mechanical devices, although very promising for their compactness, suffer from a relatively slow response ($\sim \mu$s). For cryogenic and fast operation, a device based on the electro-optic effect is therefore a much more promising solution\cite{liu_review_2015}.

To apply and control the electric field required for altering the refractive index in waveguides, doped layers can be introduced in GaAs to form junction diodes. In particular, p-i-n junctions have been widely employed in combination with QDs for controlling the emission wavelength\cite{bennett_giant_2010}, the charge state\cite{carter_quantum_2013}, and the photon coherence\cite{Kirsanske_indistinguishable_2017}. Additionally, this platform offers the opportunity to combine an electro-optic phase shifter with QDs. 
Here we demonstrate a switching circuit based on electro-optic phase shifters made of suspended GaAs waveguides. An electric field is applied to the waveguides using doped layers as electrical gates in a p-i-n junction. A $\pi$-phase shift is realized using 400-$\mu$m-long waveguides with a total applied voltage of $V_\pi=$ 2.5 V. A compact photon router is constructed by integrating two orthogonal phase shifters in a Mach-Zehnder interferometer with $(55\pm 8)$ ns response time. The device is operated with single photons from embedded In(Ga)As single quantum dots emitting at around 900 nm. By modulating the field across the junction, we observe an anti-correlated output signal with a visibility of 53\%, limited by reflections in the circuits. The single-photon nature of the emission from our QDs has been reported in a recent work performed on the same wafer\cite{Kirsanske_indistinguishable_2017}.

\begin{figure}[ht!]
\centering
\includegraphics[width=\linewidth]{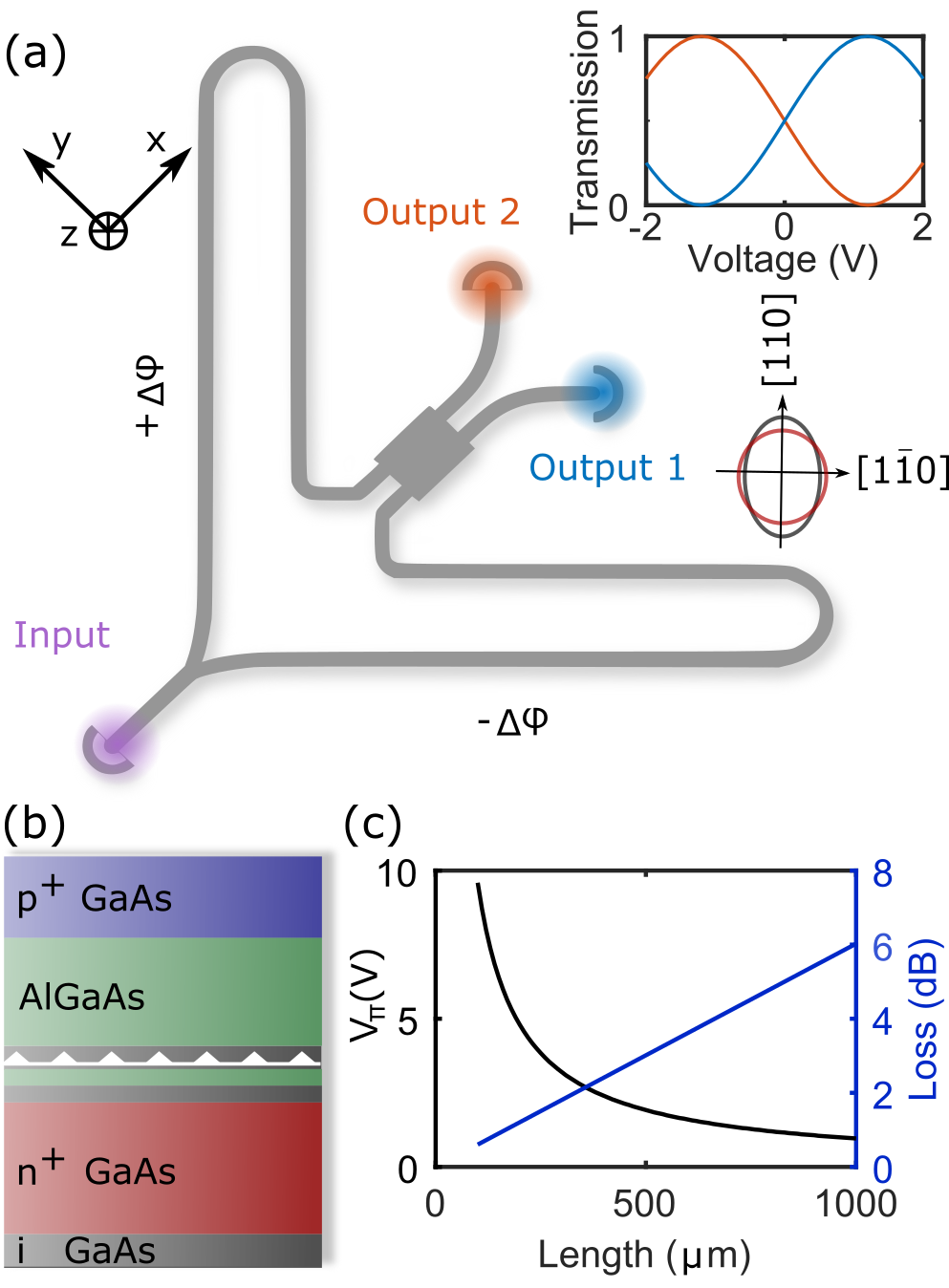}
\caption{Integrated router based on suspended electro-optic waveguides. (a) Schematic outline of the device. In the presence of an electric field in the growth direction (z), the refractive index changes along the [110] and [$1\bar{1}0$] directions as indicated by the index ellipse. The phase change results in an anti-correlated output at the two output ports as shown in the inset for an ideal loss-less device. (b) Layer structure of the electro-optic waveguides. The white triangles indicate a layer of self-assembled quantum dots. (c) Plot of the absorption loss (right axis) and switching voltage $V_{\pi}$ (V) (left axis) as a function of length of the device. A device with 400 $\mu$m-long arms requires less than 2.5 V to achieve a full switching cycle, while losses are kept below 3 dB.}
\label{fig:f1}
\end{figure}

\section{Device principles and methods}
Gallium arsenide exhibits a linear electro-optic effect along the [110] and [1$\bar{1}$0] directions when an electric field is applied along the growth direction [001]. Importantly, the effect is opposite in the two directions as a consequence of birefringence in the zincblende crystal symmetry\cite{syms_optical_1992, lee_analysis_1991}.  
The refractive index change is given by $\Delta n = \pm \frac{1}{2}n^3r_{41}E$ where $E$ is the applied electric field, $r_{41}\simeq 1.6$ pm/V is the electro-optic coefficient at a wavelength of 900 nm\cite{berseth_r41_1992}, and $n$ is the refractive index. 

The dependence on the propagation direction means that the phase shift experienced by light propagating by a distance $L$ along the [110] and $[1\bar{1}0]$ directions differs by $2\left|\Delta n\right|kL$, where $k$ is the free-space wave number. This can be exploited to double the electro-optic effect in a MZI where the two paths are orthogonal to each other as shown in Fig.~\ref{fig:f1}(a). This configuration greatly simplifies the device design since the entire circuit can be biased with the same voltage. The voltage required to achieve a $\pi$ phase difference in the two arms is given by $V_\pi = \frac{d\pi}{kL n^3 r_{41}}$, where $d$ is the distance over which the field is applied. 
The electric field is applied using a p-i-n junction integrated in the GaAs/AlGaAs heterostructure as shown in Fig.~\ref{fig:f1}(b), with a total intrinsic-region thickness of $d \sim 70$ nm. The built-in voltage $V_{b}\simeq1.5$ V readily provides an electric field of $\sim 20$ MV/m, which can be reduced or increased by operating the diode in forward or reverse bias, respectively. 

The propagation constant of the fundamental transverse electric (TE) mode in the presence of the electric field was calculated using finite-element method mode analysis (see Fig.~S1 of the Supplmentary Information). We neglect the transverse magnetic modes since the QDs only couple to TE as a result of the optical selection rules\cite{lodahl_interfacing_2015} and their placement in the center of the waveguide.      
The simulation also takes into account the free-carrier absorption introduced by the doped layers\cite{casey_absorption_1975}. Figure \ref{fig:f1}(c) shows the theoretical $V_\pi$ and insertion loss as a function of the arm length $L$. A trade-off between the accumulated phase difference and absorption determines the optimum length of the MZI. Additionally, since the low-temperature (10 K) QD emission energy in our experiment ($E_{QD}=1.37$ eV) is relatively close to the GaAs band edge ($1.52$ eV), a too large electric field will lead to an additional electroabsorption due to the Franz-Keldysh effect\cite{Franz_Keldysh_1976}. Neglecting other sources of loss, a length of 400 $\mu$m allows us to design a switch with an insertion loss as small as 2 dB and $V_\pi = 2.5$ V. Such a low switching voltage per unit length ($V_\pi L=0.1$ V$\cdot$cm) reduces the footprint of our switches by one to two orders of magnitudes compared to typical cm-long electro-optic modulators in Lithium Niobate\cite{sharapova_generation_2017} or GaAs\cite{Wang_GaAs_circuit_2014}. 

\subsection*{Device fabrication}
\begin{figure}[ht!]
\centering
\includegraphics[width=\linewidth]{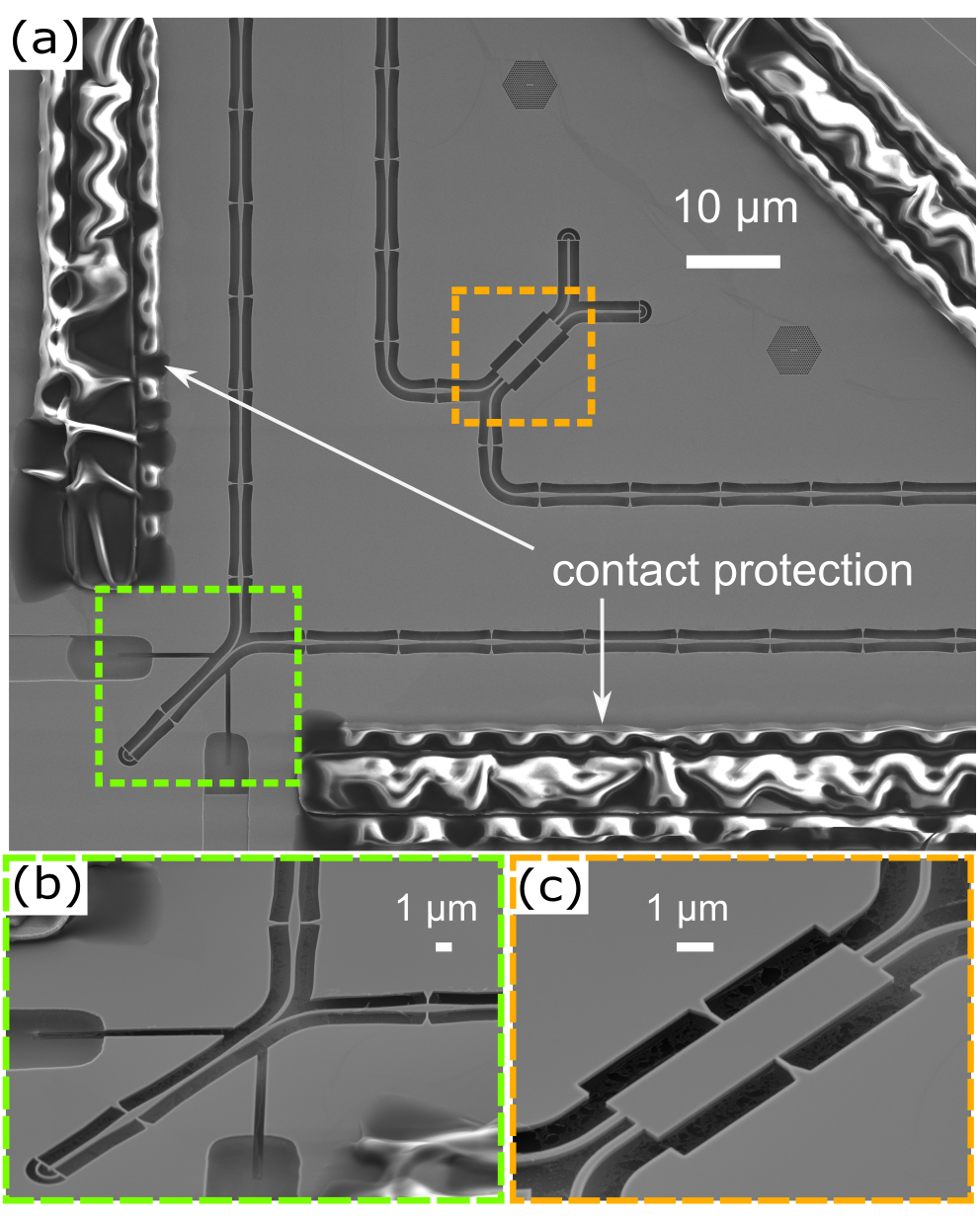}
\caption{Scanning electron micrograph (SEM) of the on-chip electro-optical router. (a) The full MZI with the two orthogonal arms. The green and yellow boxes indicate the input power splitter (Y-splitter) and 3 dB multi-mode interference (MMI) combiner, respectively. The wavy pattern on the contact protection is an artifact due to charging effects. The two arms of the MZI are suspended with 100-nm-wide tethers. (b) The fabricated Y-splitter with trenches to electrically isolate the switch area from the emitter region. (c) SEM of the output 2x2 MMI beam splitter.}
\label{fig:f1b}
\end{figure}

The device is fabricated on a [001] GaAs wafer grown by molecular beam epitaxy with the composition shown in Fig.~\ref{fig:f1}(b). A layer of self-assembled InAs QDs are grown in the middle of the membrane encapsulated in intrinsic GaAs and surrounded above and below by Al$_{0.3}$Ga$_{0.7}$As barriers to prevent carrier tunneling \cite{bennett_giant_2010}. The complete detailed wafer layout is shown in Fig.~S1(a) of the Supplementary Information. To electrically isolate each device and to expose the n-layer to metallization, two sets of trenches with different depths are etched in the GaAs membrane. An $\approx$ 100-nm-deep opening is defined by ultra-violet laser diode writing and subsequently etched by reactive ion etching (RIE) in a BCl$_3$/Ar (1:2) chemistry. Using the same processing steps, a deeper etch ($\approx$ 180 nm) is performed to create two isolated p-i-n regions, one for the voltage control of the QD and one for the electro-optic modulation of the MZI. The trenches are partially visible in figure \ref{fig:f1b}(a) and (b). An ohmic contact to the n-type GaAs is fabricated by evaporating Ni/Ge/Au/Ni/Au (5/40/60/27/100 nm) contacts followed by rapid thermal annealing at 420 $^\circ$C. For p-type contacts we use Au/Zn/Au (20/50/150 nm) and a short annealing at 380 $^\circ$C for 5 seconds.
Before the fabrication of the nanophotonic circuit, we spin-coat and pattern a layer of photo-resist (AZ1505) on top of the p-type contacts to protect them from galvanic erosion during subsequent steps. The resist is hard-baked at 185 $^\circ$C for 30 minutes. The waveguides are patterned on a 550-nm-thick ZEP520 electron beam resist layer and etched in an inductively coupled plasma RIE tool. The structures are undercut in hydrofluoric acid for 45 s, rinsed in water, and finally dried in a CO$_2$ critical-point dryer. Further details of the fabrication procedure have been presented in a previous work\cite{midolo_soft-mask_2015}. Scanning electron micrographs (SEM) of the final device are shown in Fig.~\ref{fig:f1b}. 

\subsection*{Optical and electrical characterization of the device}
The fabricated sample is attached to a copper chip carrier and wire-bonded to a printed circuit board. It is subsequently mounted on a cold-finger in a liquid-helium flow cryostat with coaxial feedthroughs and cooled to 10 K. The QDs are excited at $\lambda=808$ nm with a Ti:sapphire laser focused through a microscope objective lens with NA=0.6. The QD fluorescence is collected with the same objective and coupled into a single-mode fiber, which enables spatial filtering, thus only collecting the light scattered out by a specific grating. The two output gratings of Fig.~\ref{fig:f1b}(a) are orthogonal to each other, allowing to further suppress stray light from the other port by polarization filtering (i.e. using a half-wave plate followed by a polarization beam splitter).
The signal is dispersed by a monochromator (McPherson) into a Si detector array. 

The spectra are collected as a function of the applied bias. To properly bias the device, special care is taken to avoid heat dissipation due to unwanted currents in the device. The fabricated isolation trenches block the current between the QD diode and the switch diode, except across the waveguides in the power-splitting region (see Fig.~\ref{fig:f1b}(a)). Here, electrical power can be dissipated in the form of heat that can lead to damage or even sublimation of the material. To avoid this, the n-contacts are shortened to a common ground during the optical characterization. The sheet resistance of the p-layer is $\approx20$ times higher than the n-layer owing to the much lower mobility of holes compared to electrons in GaAs. We have verified that no damage to the waveguides occurs in this configuration but it does limit the voltage range we can investigate, as a very large voltage-difference between the two regions will cause heat dissipation. Complete isolation could be achieved by removing a tiny portion of p-layer on the waveguide region. 

\subsection*{Numerical analysis of the device}
A numerical model has been developed to understand the deviations of the fabricated device from the simplified interference model used in figure \ref{fig:f1}(a) (inset).
We use a scattering-matrix (S-matrix) formalism to describe reflection and transmission at each optical port of the various circuit components. Frequency-domain finite element method (FEM) has been used to extract the S-matrices describing the input Y-splitter of Fig.~\ref{fig:f1b}(b) and the 50/50 multi-mode combiner of Fig.~\ref{fig:f1b}(c) around the experimental wavelengths. These simulations are described in greater detail in Supplementary Information.
The matrices obtained from FEM are combined with a perfectly matched model of the interferometer arm given by:
\begin{equation}
S_{u,v} = \left(
\begin{array}{cc}
0 & \exp(-in_{u,v}(V,\lambda)k_0L) \\
\exp(-in_{u,v}(V,\lambda)k_0L) & 0
\end{array}
\right)
\end{equation}
where $n_{u,v}(V,\lambda)$ is the complex-valued effective refractive index of the TE waveguide mode for the two horizontal ($u$) and vertical ($v$) orientations and as a function of the applied voltage $V$ and wavelength $\lambda$. The effective index is extracted from a two-dimensional (2D) eigenmode analysis of the propagation constant in the waveguide (see  Fig.~S1(b) of the Supplementary Information). Here we neglect the scattering loss induced by suspension tethers which is in the order of 0.5 dB per arm according to FEM simulations. The imaginary part of the refractive index also includes the voltage-dependent absorption described by the Franz-Keldysh effect, which we model according to the phenomenological work by Stillman et al. \cite{Franz_Keldysh_1976} 

The input/output circular gratings are also modeled as two-port devices with transmission $T$ and reflection $R=30\%$ (as predicted by FEM simulations). By cascading the S-matrices of the various building blocks numerically\cite{simpson_generalized_1981}, the transmission model of the entire switch is derived as a function of the wavelength and the biasing voltage.

\section{Results}
\subsection{Routing of photons from quantum dots}
\begin{figure}[ht!]
\centering
\includegraphics[width=\linewidth]{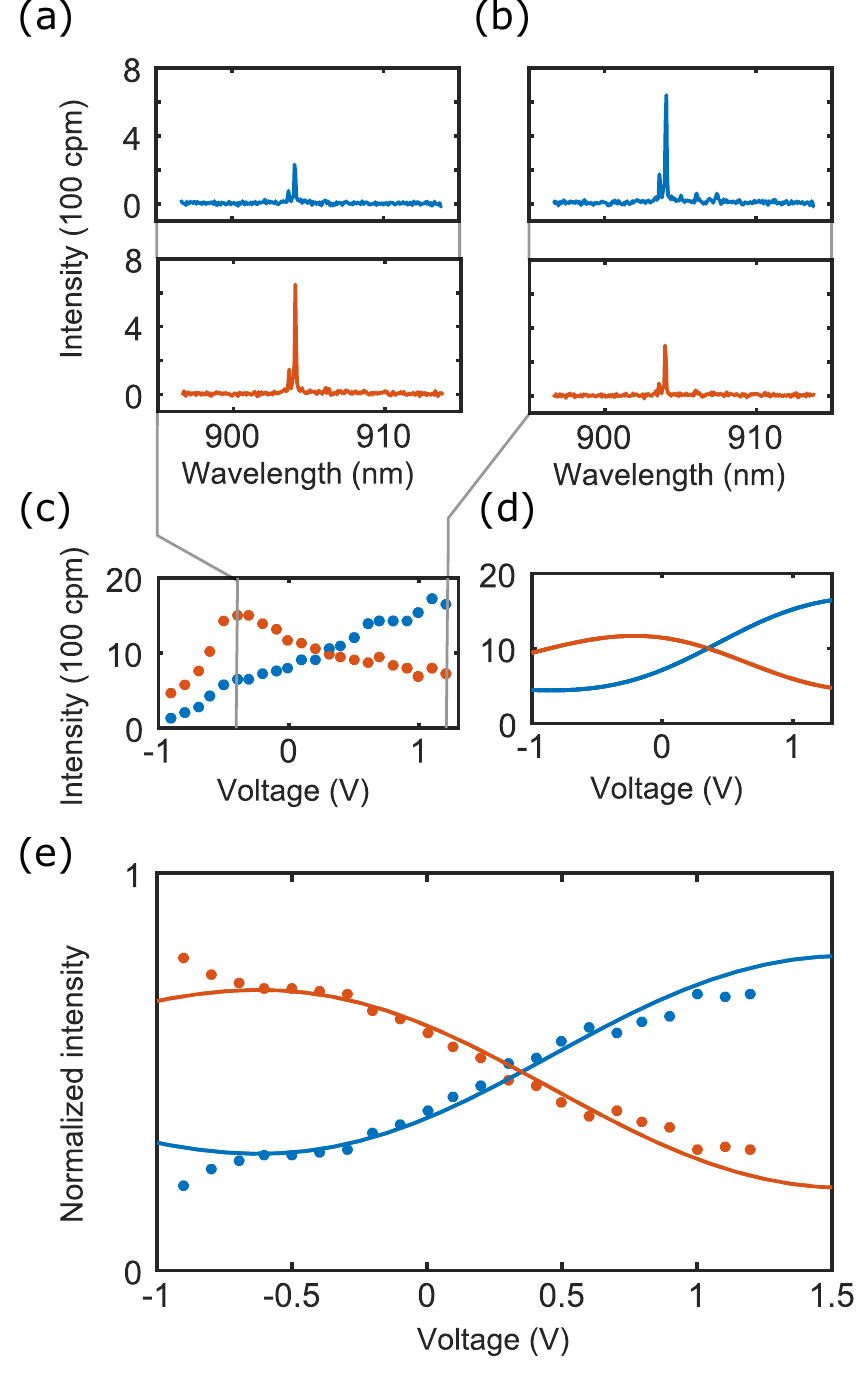}
\caption{Electro-optical switching of photons from a single quantum dot. (a,b) Single QD spectra collected at output port 1 (blue) and output port 2 (red) at -0.4V (a) and +1.2V (b). (c) Experimental data showing the integrated intensity of the QD emission as a function of the bias. The vertical lines indicate the voltages at which the spectra in (a) and (b) are recorded. (d) Theoretical predictions of the output at port 1 (blue) and port 2 (red) as a function of bias scaled to the experimental counts (details of the model are given in the Methods section). (e) Comparison between the numerical model and the experimental transmission from a quantum dot. The normalized intensity is extracted as the fraction of power emitted in one port divided by the sum of both ports.}
\label{fig:f2}
\end{figure}

Transmission measurements are performed at a temperature of 10 K using a single QD located at the input grating as an integrated light source. In order to compensate the built-in field, and thus minimise carrier tunneling from the QD, a constant bias of 1.15 V is applied in the emitter region throughout the measurement. 
The QD is excited at the input grating and the emitted photons are collected from the output gratings. The bias across the MZI region is changed in steps of 0.1 V from -1 V to 1.2 V. Figure \ref{fig:f2}(a) (\ref{fig:f2}(b)) shows the QD spectrum collected from output port 1 (blue) and output port 2 (red) when the voltage across the MZI region is -0.4 V (+1.2 V). The spectra are fitted with a Lorentzian function and the total integrated intensity as a function of the voltage across the MZI is shown in Fig.~\ref{fig:f2}(c). These data show an anti-correlation in the emission intensity between the two output ports in a range between -0.5 V and +1.2 V. At voltages below -0.5 V the intensity of the signal collected at both arms decreases with the field strength. Since the loss occurs in both arms, we attribute this effect to electro-absorption in the GaAs layers (Franz-Keldysh effect). This is confirmed by the fact that the same behavior is observed when characterizing the device with an external laser source (see Fig.~S3 of the Supplementary Information), ruling out other mechanisms such as the quenching of the QD emission due to cross-talk between p-i-n junctions. The simulated circuit transmission as a function of voltage is shown in Fig.~\ref{fig:f2}(d). 
Figure \ref{fig:f2}(e) shows the same data of Fig.~\ref{fig:f2}(c) normalized to the sum of the two output intensities and a comparison with the theoretical model (solid lines), which has been fitted to the data using only one free parameter, namely the crossing point of the two curves (at $V=0.3$ V). The MZI is therefore unbalanced at 0 V due to the presence of a built-in voltage and a slight asymmetry in the two arm lengths due to unavoidable fabrication imperfections.  
As predicted by theory, the extinction ratio is limited to $\approx$ 3.3. This limitation stems from the high reflectivity of the gratings, which introduce Fabry-P\'{e}rot (FP) resonances in the entire device. These resonances result in an additional phase shift which interferes with the ideal operation of an impedance-matched circuit. From the transmission measurements performed with a tunable laser (Fig.~S3 of the Supplementary Information), we conclude that the FP modes are not responsible for the anti-correlated output but merely add intensity fluctuations as a function of wavelength and reduce the visibility of the switching signal. In other words, the underlying physical mechanism is effectively described by the phase difference accumulated over the two arms as predicted by our device design, and not by the tuning of FP resonances. 

\subsection{Sub-microsecond response time}
\begin{figure}[ht!]
\centering
\includegraphics[width=\linewidth]{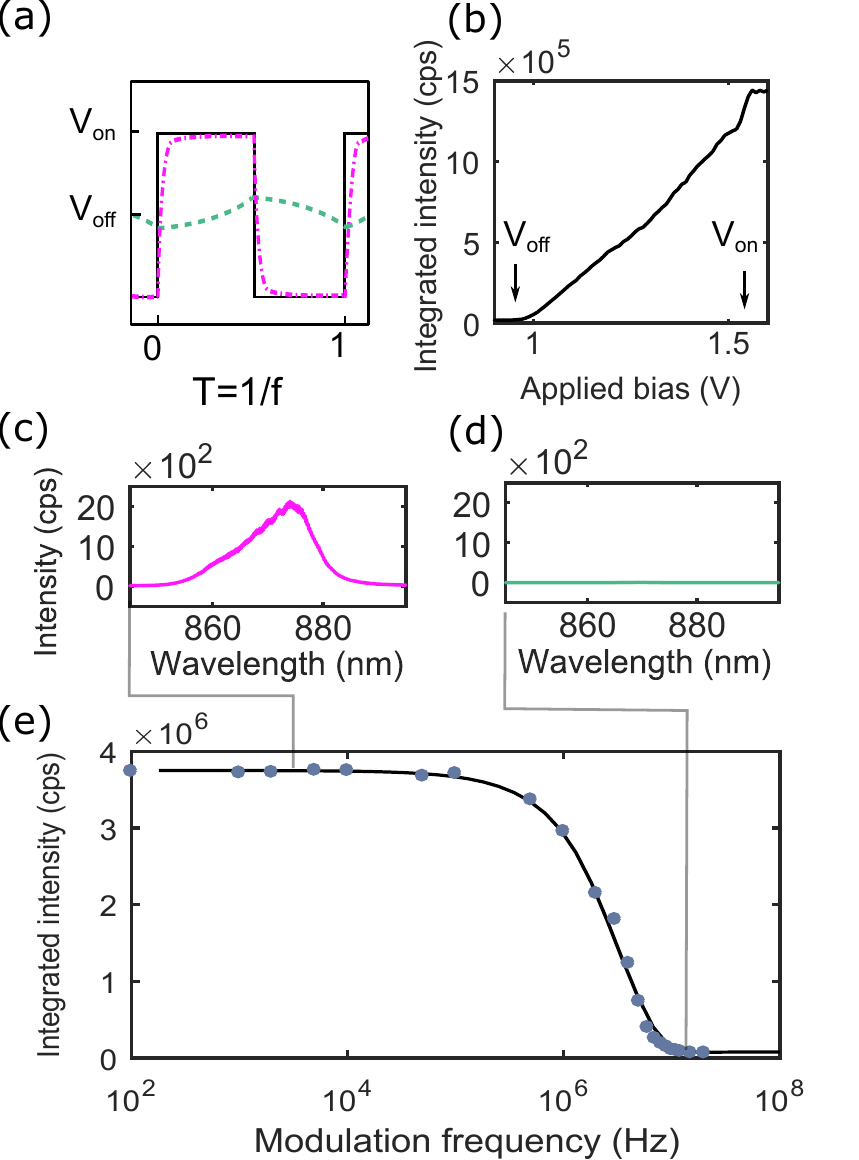}
\caption{Response time of the device. (a) A square wave voltage is applied to the sample, where $V_{\text{on}}$ and $V_{\text{off}}$ are chosen so that the sample emission is maximized at $V_{\text{on}}=1.55$ V and is completely turned off at the average value $V_{\text{off}}=0.9$ V. The purple dash-dotted (green dashed) line indicates the calculated system response below (above) the frequency response cut-off. (b) Integrated intensity from the wetting layer as a function of the applied bias. (c) The emission spectrum recorded at a modulation frequency of 10 kHz. (d) Same as (c) but at 10 MHz. (e) The integrated intensity of the wetting layer as a function of the modulation frequency of the square wave voltage (blue dots). The solid black line is the simulated response for a single-pole low-pass filter. The 3 dB cut-off is observed at around 2.8 MHz.}
\label{fig:f3}
\end{figure}
A crucial figure of merit for re-configurable quantum-photonic devices is the response time required for switching from one state to another. 
The typical timescales required vary from nanoseconds (the typical lifetime of a QD) up to several microseconds (electron spin coherence time in QDs)\cite{press_ultrafast_2010} depending on the applications.
In order to experimentally assess the response time of our switch, we perform confocal photo-luminescence measurements from the (QD) wetting layer located in the MZI waveguide while driving it with a periodic signal (Fig.~\ref{fig:f3}(a)) and record the change in the integrated emission spectrum as a function of frequency\cite{pagliano_dynamically_2014}. This allows to extract a low-pass amplitude characteristic directly without resorting to fast photodiodes or photon counters. More specifically, we measure the $RC$ constant of the circuit, where $R$ is the effective resistance arising from the metal-semiconductor contact and the doped layers (sheet resistance) and $C$ is the p-i-n junction capacitance, which includes the waveguides and the stray capacitance of the contacts. We infer the effective voltage applied to the waveguides by measuring the emission intensity dependency on a static applied bias (Fig.~\ref{fig:f3}(b)). The fact that the intensity of the wetting layer depends on the voltage stems from the variable tunneling rate of carriers as a function of the electric field across the membrane. We define a voltage $V_{\text{off}}=0.9$ V where the emission is ``switched off'' (Fig.~\ref{fig:f3}(d)) and another voltage $V_{\text{on}}=1.55$ V (flat-band condition) where the waveguide is luminescent (Fig.~\ref{fig:f3}(c)). Between these values, the intensity is, to a good approximation, linear with voltage. 
A square wave signal with variable frequency is applied to the MZI gates and adjusted so that the maximum voltage corresponds to $V_{\text{on}}$ and the average value to $V_{\text{off}}$ as shown in Fig.~\ref{fig:f3}(a). Assuming the circuit can be modeled as a low-pass filter, the junction voltage averages to $V_{\text{off}}$ when the applied signal frequency is above cut-off. Below cut-off, the wetting layer emits 50\% of the time, resulting in the spectrum of Fig.~\ref{fig:f3}(c). Figure \ref{fig:f3}(e) shows the integrated intensity extracted with this method as a function of the frequency, showing a clear low-pass characteristic. We model the response using the data from Fig.~\ref{fig:f3}(b)  and extract a time constant $RC=(55\pm8)$ ns corresponding to a 3 dB cut-off at $(2.8\pm0.5)$ MHz. The device is operated in forward bias and therefore dominated by the diffusion capacitance of the diode which is usually larger than the reverse-bias (or depletion) capacitance. 

\section{Discussion}
We have demonstrated an on-chip electro-optic phase shifter in a MZI, capable of performing controlled routing of single-QD emission between two spatially different optical modes at cryogenic temperatures. The device operates at low $V_{\pi}$ while still having a very small footprint, making it suitable for scalable quantum photonic networks. 

Further work is needed to improve the extinction ratio to the range of $10^{-4}$--$10^{-3}$ needed for boosting the efficiency in applications such as boson sampling. This could be achieved by using gratings with reduced back-scattering or adiabatic mode converters for side-coupling in fibers\cite{Kirsanske_indistinguishable_2017}. 
The overall insertion loss could be further reduced by improving the design of the p-i-n junction to minimize the overlap between the optical mode and the doped layers and by operating at longer wavelengths i.e., farther away from the GaAs absorption edge, where electro-absorption is less prominent. 
The device response time could also be further improved by reducing the active area of the MZI i.e., the stray capacitance of the diode mesa, or by further reducing the contact resistance, for example optimizing the p-type contacts or increasing the thickness of the p-doped layer.

\section{Conclusions}

The integration of a single-photon emitter and an optical router constitutes a fundamental technological step towards building reconfigurable quantum photonic devices. Although GHz switching rates would ideally be needed for tasks such as de-multiplexing, the sub-microsecond response time already makes it feasible to perform feedback operations on QDs. It could, for example, enable the realization of quantum gates based on the Duan-Kimble protocol\cite{duan_scalable_2004}, where two photons interact with a QD at different times within the coherence time. An implementation of such protocol involving QDs in chiral photonic circuits and a controlled switch has been proposed recently\cite{sollner_deterministic_2015}. Cascading multiple routers will allow building 1-to-N photon de-multiplexers\cite{lenzini_active_2016}. Harvesting the full potential of this technology requires addressing issues related to packaging and chip-to-fiber coupling. These issues have largely been resolved in telecom photonics but future work should address the transfer of these technologies to the suspended GaAs waveguide technology.

\section*{Funding Information}
Danish Council for Independent Research (Technology and Production Sciences); Villum Foundation; Quantum Innovation Center (Qubiz); European Research Council (ERC Advanced Grant ``SCALE''); DFG-TRR160; BMBF - Q.com-H 16KIS0109; DFH/UFA CDFA-05-6. 
\section*{Acknowledgments}
The authors would like to acknowledge A. Javadi and H. Thyrrestrup for technical assistance during the measurement, T. Pregnolato for assistance in fabrication, M.~C. L\"{o}bl, I. S\"{o}llner, and R.~J. Warburton for important contributions to the wafer design, and M. A. Broome and T. Schr\"{o}der for useful discussions.


\renewcommand{\thefigure}{S\arabic{figure}}
\setcounter{table}{0}
\setcounter{figure}{0}

\pagebreak
\widetext
\begin{center}
\textbf{\large Electro-optic routing of photons from single quantum dots in photonic integrated
circuits: supplementary material}
\end{center}

\begin{figure}[ht!]
\centering
\includegraphics[width=14cm]{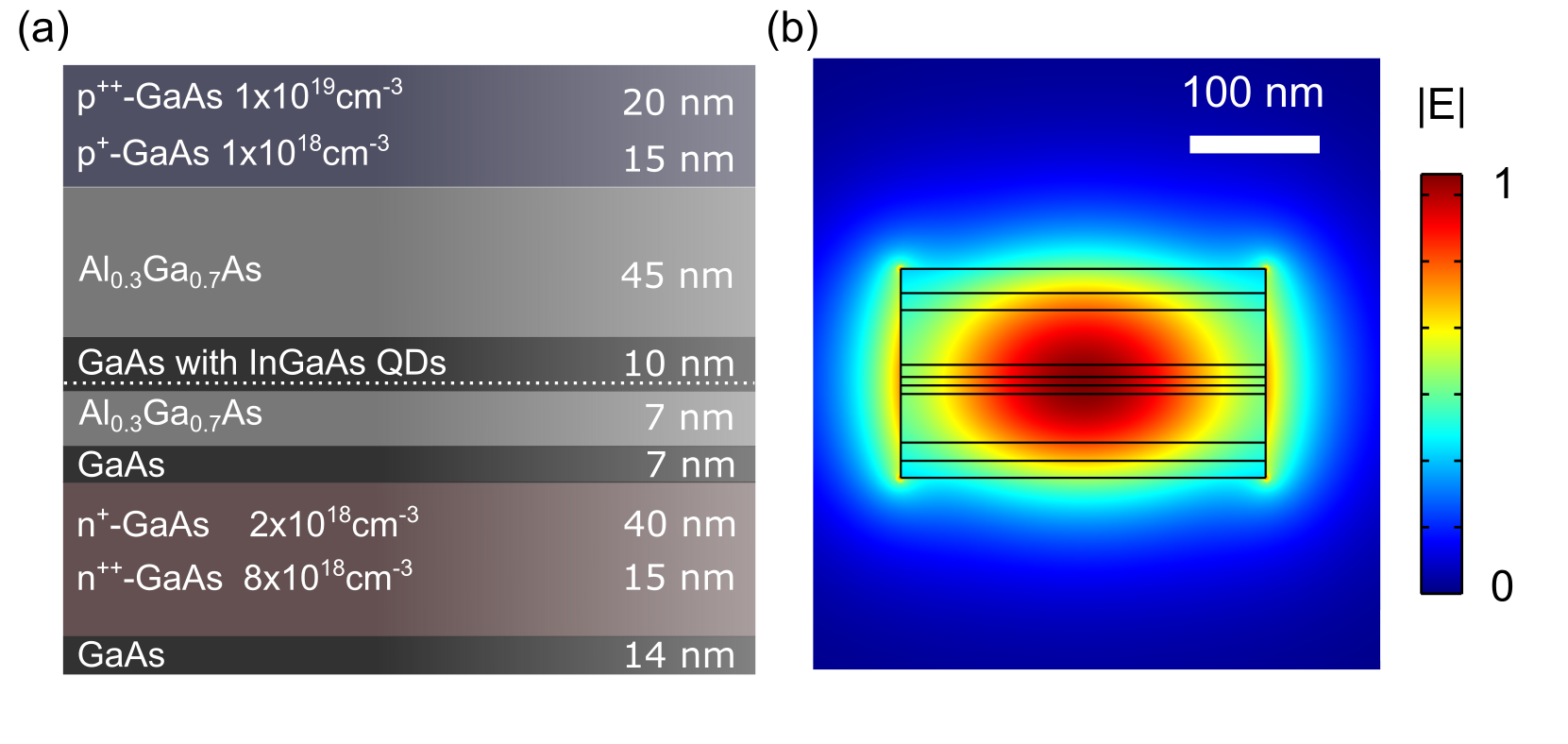}
\caption{Heterostructure layout and waveguide mode. (a) The layout of the wafer used in the present work. (b) Finite-element simulation of the fundamental transverse electric mode. The profile of the norm of the electric field is shown. The overlap of the field with the various layers defines the effective electro-optic coefficient and the magnitude of absorption. The absorption used in the model is given by free-carrier absorption $\alpha$ in the doped layers\cite{casey_absorption_1975}: $\alpha_p = 10^2$ cm$^{-1}$ for a p-doping concentration $p=10^{19}$ cm$^{-3}$ and $\alpha_n = 10$ cm$^{-1}$ for a n-doping concentration $n=5\times 10^{18}$ cm$^{-3}$. Additionally, electro-absorption is included according to the model of Stillman et al. \cite{Franz_Keldysh_1976}}
\label{fig:sf1}
\end{figure}

\begin{figure}[ht!]
\centering
\includegraphics[width=12.5cm]{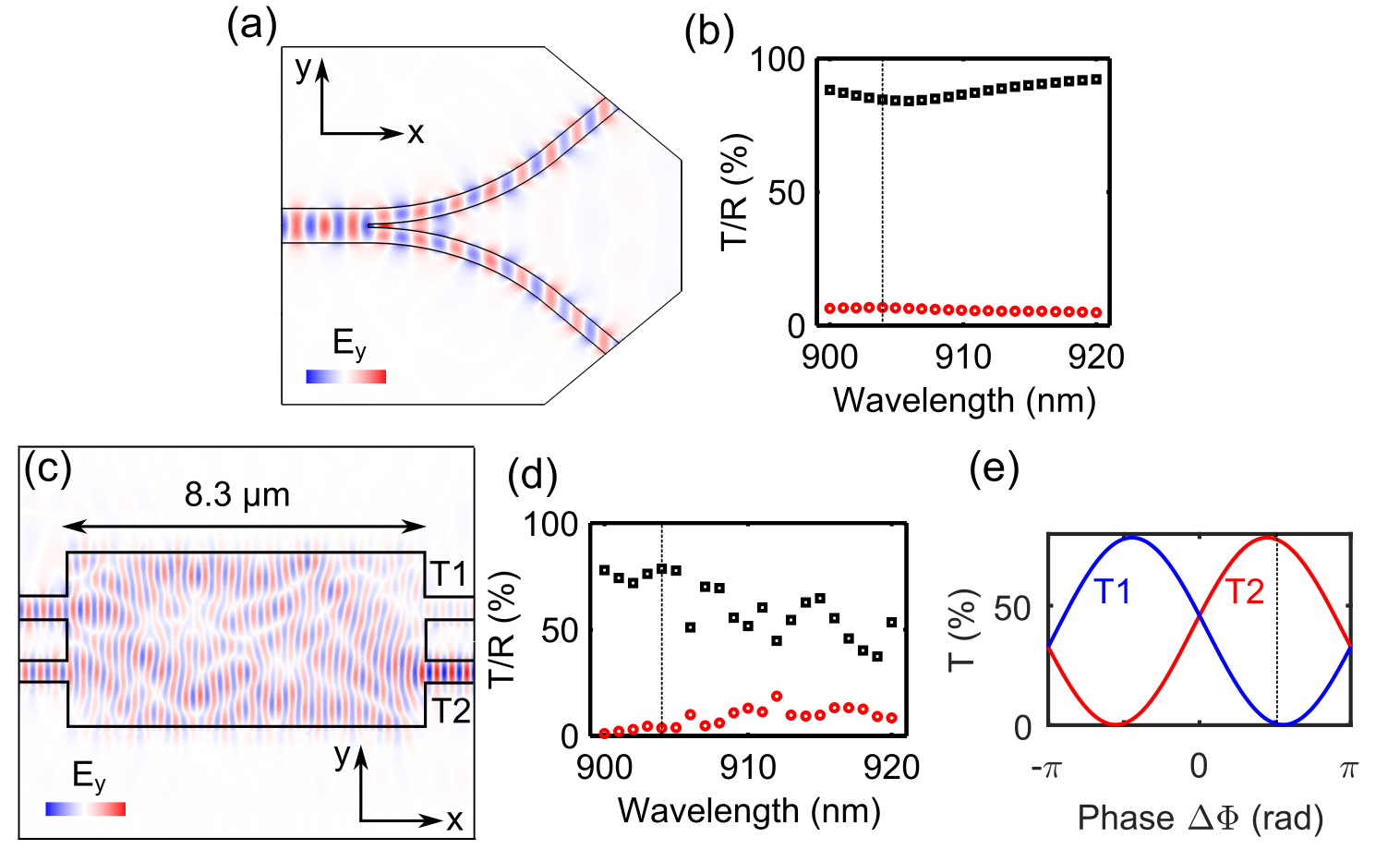}
\caption{Numerical analysis of the circuit components. (a) Finite element method (FEM) simulation of the electric field (y-component) propagation in the Y-splitter at a wavelength of 904 nm. The mirror symmetry guarantees equal power and phase at the two outputs. (b) Simulated total (sum of both ports) transmission efficiency (black squares) and reflectivity (red circles) of the Y-splitter around the quantum-dot emission wavelength. The black dotted line indicates the wavelength of our experiment  ($\lambda=904$ nm). (c) Simulated electric-field propagation in the multi-mode interference coupler (MMI) (see Fig.~1e in the main text) when launching the same power on both ports with a relative phase difference of $\pi/2$.  (d) Simulated total transmission (black squares) and reflection (red circles) in the MMI. The transmission level indicates the sum of both outputs T1 and T2 indicated in (c). (e) Transmission at the two output ports of the MMI as a function of the phase difference at the wavelength of our experiment.}
\label{fig:sf2}
\end{figure}

\begin{figure}[ht!]
\centering
\includegraphics[width=9.2cm]{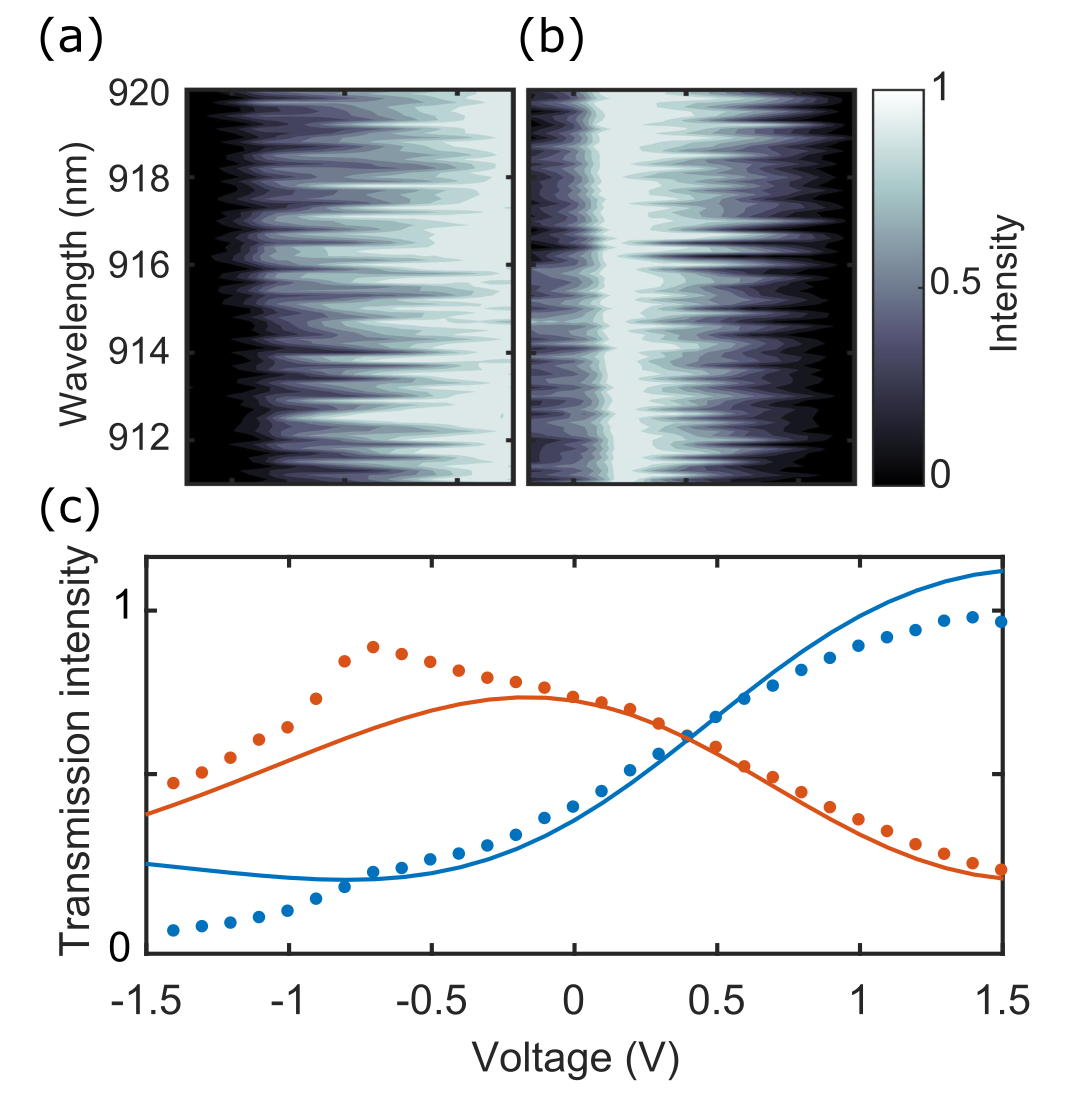}
\caption{Transmission measurements using a coherent light source. Intensity measured from (a) output port 2 (see Fig.~1a in main text) and (b) output port 1 as a function of wavelength and voltage. (c) Comparison between the theoretical and the experimental normalized transmission intensity at $\lambda=918.7$ nm. A clear drop in transmission is observed for both arms for $V<-0.7 $ V due to electro-absorption. The oscillations originate from Fabry-P\'{e}rot modes caused by reflections in the circuit.}
\label{fig:sf4}
\end{figure}

\section*{Finite element analysis of the circuit elements}
To split the input power equally, we have designed a 3 dB Y-junction splitter \cite{vlasov_active_2005}, consisting of a single-mode waveguide (240 nm wide) tapered out to a cross-section of 500 nm and split in two 240-nm-wide waveguides with a curvature radius of 3 $\mu$m (Fig.~\ref{fig:sf2}(a)). Three-dimensional finite element method (FEM) simulations show a uniform and broadband power splitting with a total insertion loss $< 0.75$ dB (sum of both output ports) and a reflection $< -11$ dB (Fig.~\ref{fig:sf2}(b)). The corresponding field distribution at a wavelength of 904 nm is shown in Fig.~\ref{fig:sf2}(c). 
To implement a 4-port 50:50 beam splitter we used the multi-mode interference (MMI) coupler shown in  Fig.~2c of the main manuscript. The device consists of a multi-mode waveguide section with a length of 8.3 $\mu$m and a width of 2 $\mu$m, supported by tethers in the middle to improve the structural stability of the suspended circuit during fabrication. The two MZI arms are merged in the multi-mode section at a distance of 1.36 $\mu$m, which is chosen to achieve paired interference leading to the formation of a self-image at the end of the multi-mode section. \cite{soldano_optical_1995}
By modifying the relative phase of the two input ports, a constructive or destructive interference is obtained at the two output ports (T1 and T2 in the figure), as expected in a 50:50 beam-splitter. The calculated field distribution (phase difference $\Delta\Phi=\pi/2$) and the total transmission (T1+T2) and reflection are shown in Figs. \ref{fig:sf2}(c) and \ref{fig:sf2}(d), respectively. The device has been optimized to reduce the insertion loss and to achieve a sufficiently balanced splitting. The simulated transmission as a function of $\Delta\Phi$ (Fig.~\ref{fig:sf2}(e)) shows a slight shift between the two transmission outputs. We neglect this effect when modeling the entire device, as we do not see a significant deviation from an idealized model of a MMI coupler.

\section*{Characterization of the device transmission}
To verify that the transmission across the circuit is independent from the biasing of the quantum dots, the device is tested with an external source, specifically with a continuously tunable continuous wave laser (Toptica CTL). 
The voltage was swept from -1.5 V to +1.5 V and the intensity at the two output-ports was measured from the image on a charge-coupled device (CCD) camera. To ensure that the observed transmission refers to the correct polarization in the waveguide (transverse electric), the input polarization has been carefully aligned to the input grating. 
Figs.~\ref{fig:sf4}(a) and \ref{fig:sf4}(b) show the intensity measured at the output ports as a function of voltage and wavelength.
For each wavelength the data are normalized using the gray scale image normalization in order to correct for any power fluctuations of the laser. The expected anti-correlated behavior at the output gratings was confirmed for a wavelength range of 911-920 nm, which confirms that the index-modulation caused by the voltage on the Fabry-P\'{e}rot fringes in the circuit is not the origin of the observed switching characteristic. If this were the case, we would observe a more narrow-band emission at the output ports. 
The experimental data are also compared to the predictions of the theoretical model discussed in the Methods section of the main manuscript (Fig.~\ref{fig:sf4}(c)). As for the case of QD transmission, we also observe quenching of the transmission at negative bias voltages around -0.7 V due to electro-absorption.

\end{document}